\documentclass[aps,prl,twocolumn,superscriptaddress,showpacs,showkeys]{revtex4-2}

\usepackage{color}
\usepackage{graphicx}
\usepackage{dcolumn}
\usepackage{bm}
\usepackage{float}
\usepackage[mathlines]{lineno}
\usepackage{tabularx}
\usepackage[colorlinks,linkcolor=blue,anchorcolor=blue,citecolor=blue,urlcolor=blue,hyperindex,CJKbookmarks]{hyperref}
\usepackage{footnote}
\usepackage{amsmath}
\usepackage{txfonts}
\usepackage{mathptmx}
\usepackage{ragged2e}
\usepackage{booktabs,makecell, multirow, tabularx}
\usepackage{multirow}
\usepackage{braket}
\usepackage{gensymb}
\usepackage{array}

\preprint{APS/123-QED}

\begin{document}

\title{High temperature superconductivity in Li$_2$AuH$_6$ mediated by strong electron-phonon coupling under ambient pressure}

\author{Zhenfeng Ouyang}\thanks{These authors contributed equally to this work.}\affiliation{School of Physics and Beijing Key Laboratory of Opto-electronic Functional Materials $\&$ Micro-nano Devices, Renmin University of China, Beijing 100872, China}\affiliation{Key Laboratory of Quantum State Construction and Manipulation (Ministry of Education), Renmin University of China, Beijing 100872, China}

\author{Bo-Wen Yao}\thanks{These authors contributed equally to this work.}\affiliation{School of Physics and Beijing Key Laboratory of Opto-electronic Functional Materials $\&$ Micro-nano Devices, Renmin University of China, Beijing 100872, China}\affiliation{Key Laboratory of Quantum State Construction and Manipulation (Ministry of Education), Renmin University of China, Beijing 100872, China}

\author{Xiao-Qi Han}\thanks{These authors contributed equally to this work.}\affiliation{School of Physics and Beijing Key Laboratory of Opto-electronic Functional Materials $\&$ Micro-nano Devices, Renmin University of China, Beijing 100872, China}\affiliation{Key Laboratory of Quantum State Construction and Manipulation (Ministry of Education), Renmin University of China, Beijing 100872, China}

\author{Peng-Jie Guo}\affiliation{School of Physics and Beijing Key Laboratory of Opto-electronic Functional Materials $\&$ Micro-nano Devices, Renmin University of China, Beijing 100872, China}\affiliation{Key Laboratory of Quantum State Construction and Manipulation (Ministry of Education), Renmin University of China, Beijing 100872, China}

\author{Ze-Feng Gao}\email{zfgao@ruc.edu.cn}\affiliation{School of Physics and Beijing Key Laboratory of Opto-electronic Functional Materials $\&$ Micro-nano Devices, Renmin University of China, Beijing 100872, China}\affiliation{Key Laboratory of Quantum State Construction and Manipulation (Ministry of Education), Renmin University of China, Beijing 100872, China}

\author{Zhong-Yi Lu}\email{zlu@ruc.edu.cn}\affiliation{School of Physics and Beijing Key Laboratory of Opto-electronic Functional Materials $\&$ Micro-nano Devices, Renmin University of China, Beijing 100872, China}\affiliation{Key Laboratory of Quantum State Construction and Manipulation (Ministry of Education), Renmin University of China, Beijing 100872, China}\affiliation{Hefei National Laboratory, Hefei 230088, China}

\date{\today}

\begin{abstract}
We used our developed AI search engine~(InvDesFlow) to perform extensive investigations regarding ambient stable superconducting hydrides. A cubic structure Li$_2$AuH$_6$ with Au-H octahedral motifs is identified to be a candidate. After performing thermodynamical analysis, we provide a feasible route to experimentally synthesize this material via the known LiAu and LiH compounds under ambient pressure. The further first-principles calculations suggest that Li$_2$AuH$_6$ shows a high superconducting transition temperature ($T_c$) $\sim$ 140 K under ambient pressure. The H-1$s$ electrons strongly couple with phonon modes of vibrations of Au-H octahedrons as well as vibrations of Li atoms, where the latter is not taken seriously in other previously similar cases. Hence, different from previous claims of searching metallic covalent bonds to find high-$T_c$ superconductors, we emphasize here the importance of those phonon modes with strong electron-phonon coupling (EPC). And we suggest that one can intercalate atoms into binary or ternary hydrides to introduce more potential phonon modes with strong EPC, which is an effective approach to find high-$T_c$ superconductors within multicomponent compounds.
\end{abstract}


\maketitle

\textit{Introduction.}
Searching superconducting materials with high transition temperature ($T_c$) is a hot issue since the discovery of superconductivity~\cite{onnes1911comm}. As the lightest element, high-$T_c$ superconductivity was anticipated to realize in metallic hydrogen~\cite{PhysRevLett.21.1748}. However, solid hydrogen is too difficult to stabilize and show superconductivity under ambient pressure. Consequently, researchers are shifting their focus toward hydrogen-rich compounds under high-pressure conditions. In 2014, H$_3$S was theoretically predicted to exhibit superconductivity under high pressure~\cite{10.1063/1.4874158, WOS:000344759000003}, which was experimentally confirmed~\cite{WOS:000360594100027}. With development of high-pressure technique, various hydrogen-rich superconductors with high-$T_c$ were synthesized, for example, LaH$_{10}$~\cite{WOS:000468844100045, PhysRevLett.122.027001}, YH$_9$~\cite{WOS:000404576100051, PhysRevLett.119.107001}, and CaH$_6$~\cite{doi:10.1073/pnas.1118168109}, which makes high-pressure hydride become a very promising candidate for room-temperature superconductivity. 

On the other hand, extremely high pressure brings challenge to experimental synthesis as well as characterization of sample. Applying chemical pressure by introducing heavy atoms is an effective approach, which may decrease the required pressure. And based on this, lots of binary rare-earth hydrides ReH$_n$ (Re = Ce, Pr, Nd, Eu, and Th) were theoretically proposed and successively synthesized~\cite{PhysRevLett.127.117001, PhysRevLett.119.107001,WOS:000522556600017,WOS:000478014000003,WOS:000488484400007,doi:10.1126/sciadv.aax6849,doi:10.1021/jacs.9b10439,PhysRevResearch.3.043107} but most of them still require pressure $\sim$ 100 GPa. 

In recent years, ternary hydride gets more attentions due to the enlarged phase space and enhanced flexibility of manipulation. Some superconducting ternary hydrides were theoretically proposed, for example, H$_6$SX (X = Cl, Br)~\cite{PhysRevB.105.L180508}, LaRH$_8$ (R = B, Be)~\cite{PhysRevB.104.L020511,PhysRevLett.128.047001,PhysRevB.104.134501}, Li$_2$MgH$_{16}$~\cite{PhysRevLett.123.097001}, KB$_2$H$_8$~\cite{PhysRevB.104.L100504}, and CsBH$_5$~\cite{PhysRevB.107.L180501}. Among them, the strong electron-phonon coupling (EPC) driven by metallic B-H $\sigma$-bond and related phonon in KB$_2$H$_8$ lead a superconducting transition with $T_c$ $\sim$ 134 K under 12 GPa. Furthermore, using chemical doping to modulate virtual high-pressure effect, CsBH$_5$ exhibit superconductivity with $T_c$ $\sim$ 98 K under near-ambient 1 GPa pressure. In addition, ternary hydrides LaBeH$_8$~\cite{PhysRevLett.130.266001} ($T_c$ $\sim$ 110 K under 80 GPa), LaB$_2$H$_8$~\cite{Song2024} ($T_c$ $\sim$ 106 K under 90 GPa), and La-Ce-H system~\cite{Chen2023} ($T_c$ $\sim$ 176 K under 100 GPa) were experimentally reported. Both of these theoretical and experimental works suggest that ternary hydride is expected to exhibit high-$T_c$ superconductivity under relatively low even ambient pressure.

Very recent, K. Dolui $et al$., performed $ab$ $initio$ random structure search at 1 GPa and predicted various potential superconducting ternary hydrides A$_a$B$_b$H$_c$~\cite{PhysRevLett.132.166001}. A metastable cubic Mg$_2$IrH$_6$ with predicted transition $T_c$ $\sim$ 160 K was revealed, which may be synthesized via a high-pressure route from insulating Mg$_2$IrH$_7$. Another group performed a machine-learning accelerated high-throughput investigation regarding Mg$_2$XH$_6$ (X = Rh, Ir, Pd, and Pt) compounds under ambient pressure, which suggests that superconducting $T_c$ are in range of 45 - 80 K and the $T_c$ of Pt compound may be enhanced above 100 K via doping electrons~\cite{Sanna2024}. High-throughput screening calculations of EPC in X$_2$MH$_6$ (X = Li , Na, Mg, Al, K, Ca, Ga, Rb, Sr, and In; M are transition metals) further predicted more compounds with superconducting $T_c$ exceeding 50 K under ambient pressure, which enlarges this superconducting ternary 216-type family~\cite{ZHENG2024101374}.  

Techniques from machine learning and data science are increasingly being employed to address problems in materials science~\cite{han2024aidriveninversedesignmaterials}. In particular, generative modeling methods based on autoencoder architectures and diffusion~\cite{han2024invdesflowaisearchengine,gao2025nsr} have been employed to generate crystal structures.
In this letter, we performed vast investigations regarding ternary hydrides through our developed AI search engine, namely InvDesFlow~\cite{han2024invdesflowaisearchengine}, which integrates generative AI and several graph neural networks for the discovery of high-$T_c$ superconductors. See the supplementary materials~\cite{SM} for the detail of the mentioned AI search engine (see also references [2, 9] therein). The specific setting of the AI model can be found in the source code~\cite{code}. A high-$T_c$ superconductor Li$_2$AuH$_6$ that is isostructural to Mg$_2$IrH$_6$ was identified. Combined with the first-principles density-functional theory and superconducting EPC calculations, Li$_2$AuH$_6$ shows a high-$T_c$ $\sim$ 140 K under ambient pressure. Thermodynamic study suggests that Li$_2$AuH$_6$ is a metastable phase and may be experimentally synthesized via the known LiH and LiAu compounds. Detailed EPC analysis indicates that the H-1$s$ electrons couple with special phonon modes then drive strong Cooper pairing. Different from previous reports that emphasized the vibrations of H octahedrons, we find that the vibrations of Li atoms also contribute strong EPC, even stronger than that of vibrations of H octahedrons. Our work reveals a ternary hydride Li$_2$AuH$_6$ with superconducting $T_c$ $\sim$ 140 K under ambient pressure and suggests that EPC can be enhanced by introducing more strong coupling modes.

\textit{Methods.}
In this work, the DFT calculations were performed by the first-principles package QUANTUM-ESPRESSO~\cite{Giannozzi_2009}. The generalized gradient approximation of the Perdew-Burke-Ernzerhof formula~\cite{PhysRevLett.77.3865} for the exchange and correlation functional was chosen. The optimized norm-conserving Vanderbilt pseudopotential was used~\cite{PhysRevB.88.085117}. The kinetic energy cutoff and the charge density cutoff were respectively set to be 80 Ry and 320 Ry. The charge densities were self-consistently calculated on an unshifted 16$\times$16$\times$16 $\bf{k}$-point mesh grid with a Methfessel-Paxton smearing of 0.02 Ry~\cite{PhysRevB.40.3616}. And the phonon were calculated on a 4$\times$4$\times$4 $\bf{q}$-point mesh grid based on the density-functional perturbation theory~\cite{RevModPhys.73.515}.   

The Wannier interpolation technique was used to calculate electron-phonon coupling superconductivity with electron-phonon Wannier (EPW) codes~\cite{PONCE2016116}. The maximally localized Wannier functions (MLWFs)~\cite{Pizzi_2020} were constructed on a 4$\times$4$\times$4 $\bf{k}$-point mesh grid. The Au-$5d$, and H-$1s$ orbitals were projected. The convergent EPC constant $\lambda$ was extensively carried out through fine electron (48$\times$48$\times$48) and phonon (16$\times$16$\times$16) grid and the dirac $\delta$ functions for electrons and phonon were smeared out by a Gaussian function with widths of 90 meV and 0.5 meV, respectively. The anisotropic Eliashberg equations~\cite{PONCE2016116, CHOI200366, PhysRevB.87.024505} were solved on a fine electron grid of 48$\times$48$\times$48 points. The sum over Matsubara frequencies was truncated with $\omega_{c}$ = 1.7 eV, about 10 times that of the highest phonon frequency.

According to the Migdal-Eliahsberg theory~\cite{RevModPhys.89.015003}, the mode- and wavevecter-dependent coupling $\lambda_{\textbf{q}\nu}$ reads
\begin{align}
\lambda_{\textbf{q}\nu} = \frac{2}{{\hbar}N(0)N_{\textbf{k}}} \sum_{nm{\textbf{k}}} \frac{1}{{\omega}_{{\textbf{q}\nu}}} {\vert {g^{nm}_{{\textbf{k},{\textbf{q}}}\nu}} \vert}^2 \delta(\epsilon^{n}_{\textbf{k}}) \delta(\epsilon^{m}_{\textbf{k+q}}),
\end{align}
where $N$(0) is the density of states (DOS) of electrons at the Fermi level. $N_{\textbf{k}}$ is the total numbers of {\textbf{k}} points in  the fine {\textbf{k}}-mesh. $\omega_{{\textbf{q}}\nu}$ is the phonon frequency and $g^{nm}_{{\textbf{k},{\textbf{q}}}\nu}$ is the EPC matrix element. ($n$, $m$) and $\nu$ denote the indices of bands and phonon mode, respectively. $\epsilon^{n}_{\textbf{k}}$ and $\epsilon^{m}_{\textbf{k+q}}$ are the band eigenvalues with respect to the Fermi level.
By the summation of $\lambda_{{\textbf{q}}\nu}$ over the first Brillouin zone, or the intergration of the Eliahberg spectral function $\alpha^2F(\omega)$, the EPC constant $\lambda$ was determined 
\begin{align}
\lambda = \frac{1}{N_{\textbf{q}}} \sum_{{\textbf{q}}\nu} \lambda_{\textbf{q}\nu} = 2 \int{\frac{\alpha^2F(\omega)}{\omega}} d\omega,
\end{align}
where $N_\textbf{q}$ represents the total number of \textbf{q} points in the fine \textbf{q}-mesh. The Eliashberg spectral function $\alpha^2F(\omega)$ was calculated by
\begin{align}
\alpha^2F(\omega) = \frac{1}{2N_{\textbf{q}}} \sum_{\textbf{q}\nu} \lambda_{\textbf{q}\nu}\omega_{\textbf{q}\nu}\delta(\omega - \omega_{{\textbf{q}}\nu}).
\end{align}

\textit{Results and Analysis.}
\begin{figure}[thb]
\centering
\includegraphics[width=8.6cm]{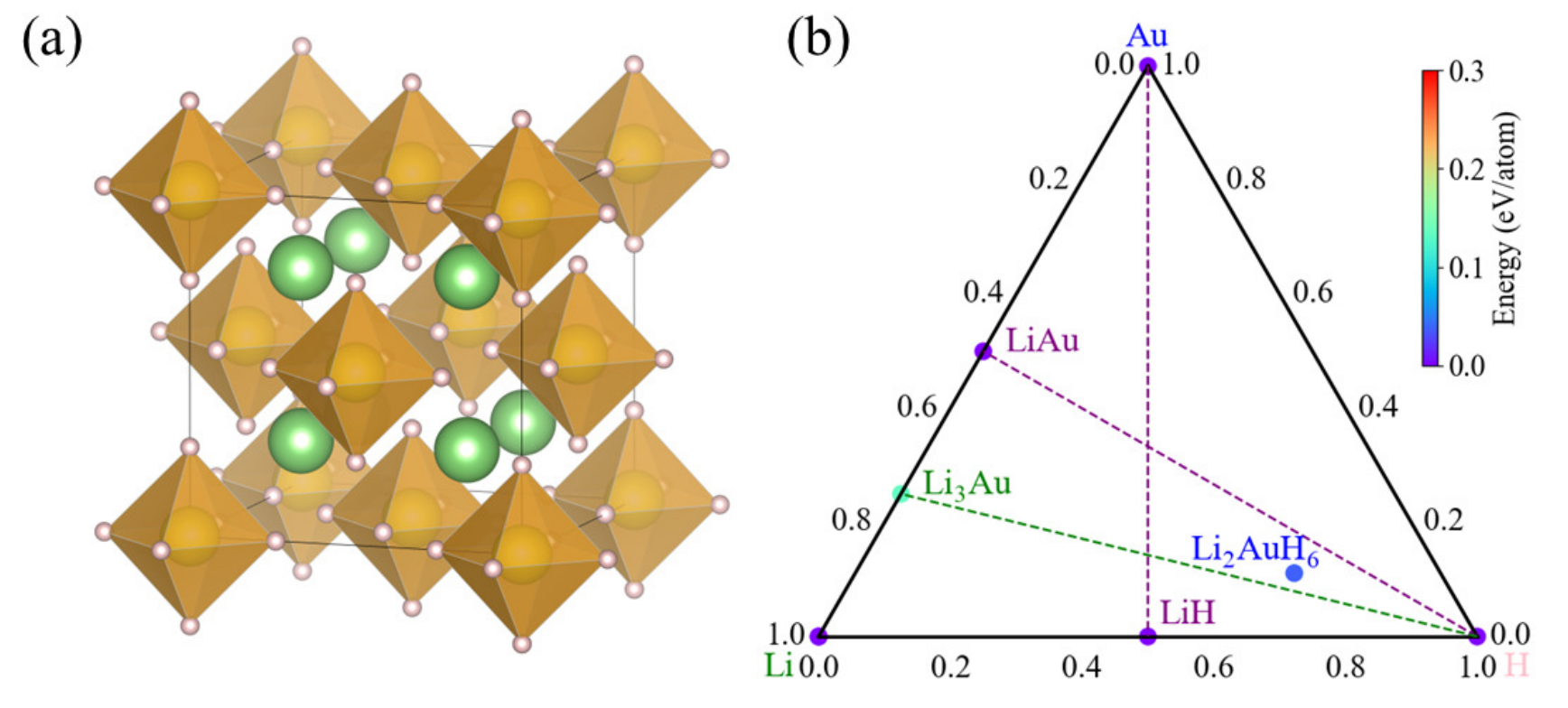}
\caption{(a) Crystal structure of Li$_2$AuH$_6$. The green, yellow, and pink balls denote the Li, Au, and H atoms, respectively. And the Au-H octahedrons are presented by golden surfaces. (b) Ternary convex hulls of Li-Au-H systems. }
\label{fig:structure}
\end{figure}

Figure~\ref{fig:structure}(a) shows the crystal structure of Li$_2$AuH$_6$. Li atoms are intercalated into the interstitial position between Au-H octahedrons and occupy the Wyckoff sites $8c$ (0.25, 0.25, 0.25). Many similar 216-type structures were theoretically proposed in previous works while a considerable number of them are energetically unfavorable under ambient pressure, which implies difficulty for experimental synthesis. Hence, we first focus the thermodynamic property of Li$_2$AuH$_6$. As shown in Fig.~\ref{fig:structure}(b), there are two possible approaches that may synthesize Li$_2$AuH$_6$ from Li-Au alloy. Our calculations suggest that LiAu possesses a lower formation energy $\sim$ $-$0.54 eV/atom at ambient pressure than Li$_3$Au ($-$0.41 eV/atom), which is consistent to a previous report~\cite{doi:10.1021/jacs.5b11768}. The formation energy is defined as
\begin{align}
{\Delta}E(\text{Li}{_m}\text{Au}{_n}) = \frac{E(\text{Li}{_m}\text{Au}{_n}) - mE(\text{Li}) - nE(\text{Au})}{m + n}.
\end{align}
Hence, we propose a feasible route 6LiH + 5Au $\rightarrow$ Li$_2$AuH$_6$ + 4LiAu. The calculated formation energy of right side of equation is only $\sim$ 38 meV/atom higher than that of left side. As for such a solid metastable phase, the reaction kinetics under ambient condition may be slow, hence catalysts may be used. In addition, metastable structure may decompose or transform into other phases during synthesis, which suggests that rapid quenching or inert gas atmospheres or encapsulation may be needed. In a word, according to a survey~\cite{Sun2016}, the lower formation energy than a typical value of 80 meV/atom implies a good experimental synthesizability. These results suggest that Li$_2$AuH$_6$ is a metastable phase that may be experimentally synthesized.

\begin{figure}[thb]
\centering
\includegraphics[width=8.6cm]{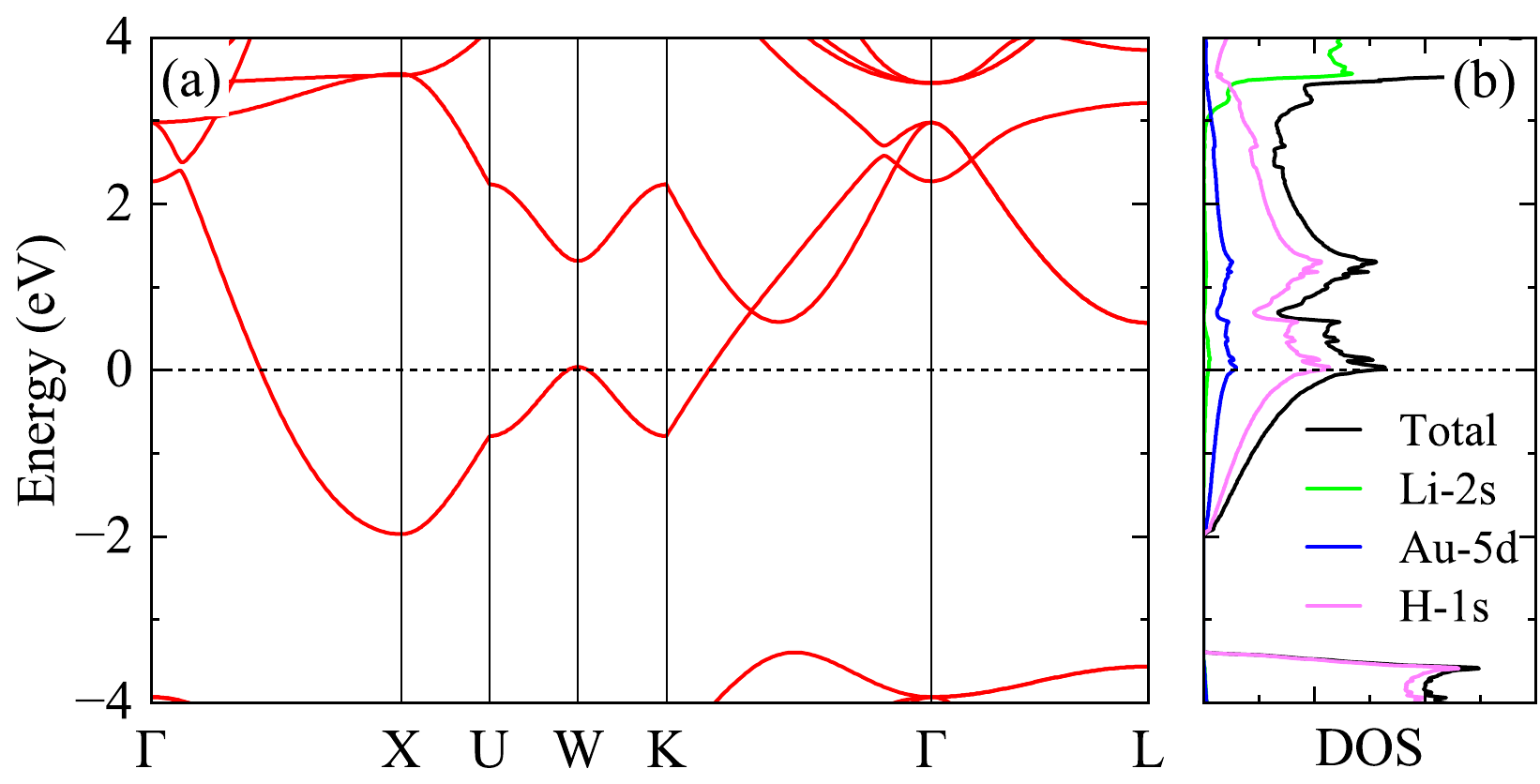}
\includegraphics[width=8.6cm]{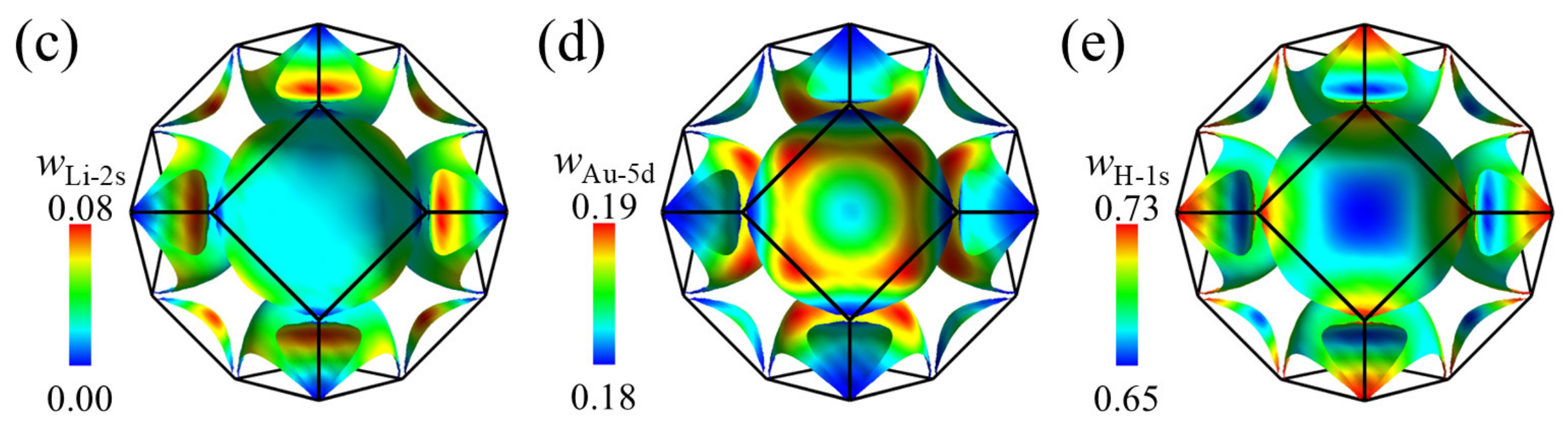}
\caption{Electronic structure of Li$_2$AuH$_6$ for (a) band structure and (b) atomic orbital-projected density of states (PDOS). The Fermi level is set to zero. (c-e) The Fermi surfaces with orbital weight of the Li-2$s$, Au-5$d$, and H-1$s$ orbitals, respectively.}
\label{fig:band}
\end{figure}

Figure~\ref{fig:band}(a) shows that Li$_2$AuH$_6$ is a metal with one band crossing the Fermi level. And the calculated PDOS in Fig.~\ref{fig:band}(b) suggests that electrons from Au-H octahedrons contribute almost all the electronic states around the Fermi level. A van Hove singularity is found at the $W $ point, which contributes a high peak of PDOS at the Fermi level. In Fig.~\ref{fig:band}(c-e), we show the Fermi surfaces with orbital weight. The contribution of Li-2$s$ orbital is negligible due to weak electronegativity of Li. And relatively low fraction of Au-5$d$ orbitals lead an anisotropic distribution of electronic states upon the Fermi surfaces. For example, in the hemispherical Fermi surfaces, the H-1$s$ orbital possesses significantly different weights around the $W$ point and at the bottom of hemisphere.

\begin{figure}[thb] 
\centering
\includegraphics[width=8.6cm]{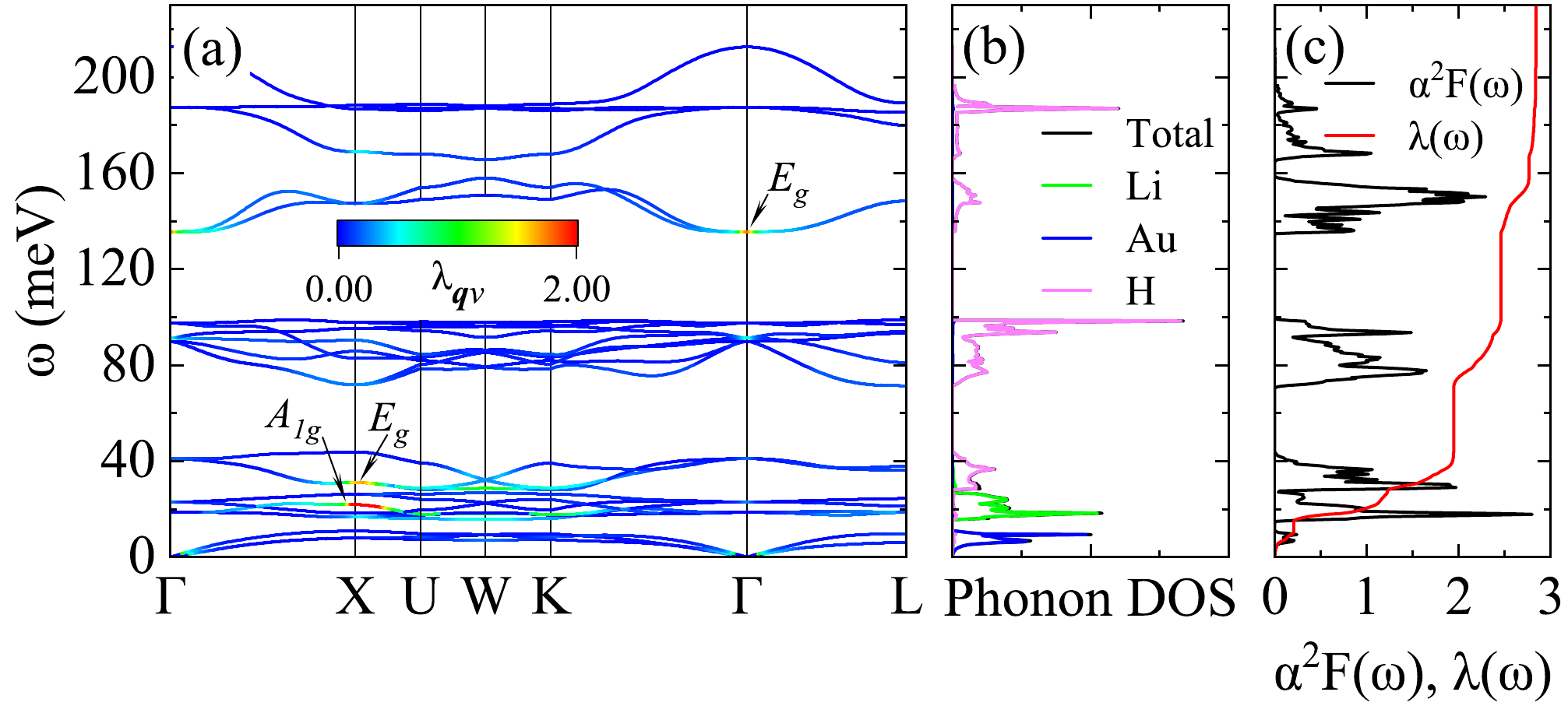}
\includegraphics[width=8.6cm]{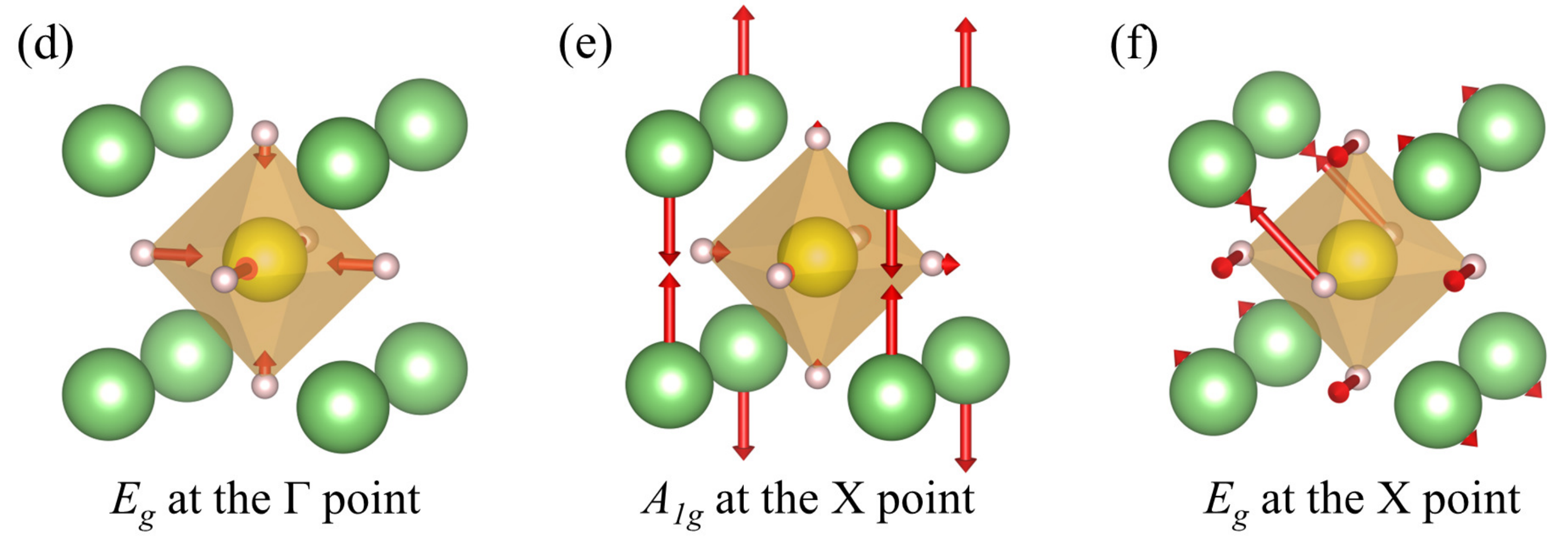}
\caption{(a) Phonon spectrum with a color representation of $\lambda_{\bf{q}\nu}$ under ambient pressure. (b) Total and projected phonon DOS. (c) Eliashberg spectral function $\alpha^2F(\omega)$ and accumulated $\lambda(\omega)$. The graduation of $\alpha^2F(\omega)$ is omitted for clarify. (d-f) The vibration modes for $E_g$ at the $\Gamma$ point, $A_{1g}$ at the $X$ point, and $E_g$ at the $X$ point, respectively.}
\label{fig:phonon}
\end{figure}

Next, we investigate the phonon and EPC of Li$_2$AuH$_6$ under ambient pressure. As shown in Fig.~\ref{fig:phonon}(a), no trace of imaginary phonon mode is found under ambient, which suggests that Li$_2$AuH$_6$ is dynamically stable. We also study some other same structure where the Au atom is respectively replaced by Pd, Ag, and Pt. We find that Au-H octahedron shows unique stability under ambient pressure, while the same case of Pd and Ag show a maximum imaginary phonon $\sim$ $-$3.2 and $-$19 meV, respectively and Li$_2$PtH$_6$ is a band insulator. As shown in Fig.~\ref{fig:phonon}(b), phonon spectrum of Li$_2$AuH$_6$ can be divided into three regions and two obvious energy gaps are observed. Specifically, high-frequency region above 120 meV and medium-frequency region ranging from 60 to 100 meV are totally contributed by vibrations of H atoms. Acoustic phonons and relatively low-frequency optical branches below 50 meV are respectively contributed by the vibrations of Au and Li mixed with H.

Furthermore, our EPC calculations suggest that the EPC constant $\lambda$ of  Li$_2$AuH$_6$ is 2.84 and there are three phonon modes that contribute strong EPC, they are $E_g$ mode $\sim$ 140 meV at the $\Gamma$ point, $A_{1g}$ mode $\sim$ 20 meV at the $X$ point, and $E_g$ mode $\sim$ 30 meV at the $X$ point, which are labeled in Fig.~\ref{fig:phonon}(a). As shown in Fig.~\ref{fig:phonon}(d), the $E_g$ mode at the $\Gamma$ point is a breathing mode of Au-H octahedron. This vibration mode may cause strong changes of distribution of charge density, hence  provide strong EPC. And this has been mentioned in the studies of other 216-type structure containing H octahedrons. However, the $A_{1g}$ and $E_g$ modes at the $X$ point are also identified to induce strong EPC in Li$_2$AuH$_6$. 
These two modes contain vibrations of Li atoms besides those of Au-H octahedrons and the corresponding vibration patterns are exhibited in Fig.~\ref{fig:phonon}(e) and (f). The accumulated $\lambda(\omega)$ suggests that the contribution of low-frequency region below 30 meV mainly dominated by these two modes reaches $\sim$ 70\% of total $\lambda$, which is significantly larger than that of the breathing mode of Au-H octahedrons. These results suggest that searching vibration mode with strong EPC would be an effective approach to discover high-$T_c$ conventional superconductors.

\begin{figure}[t]
\centering
\includegraphics[width=8.6cm]{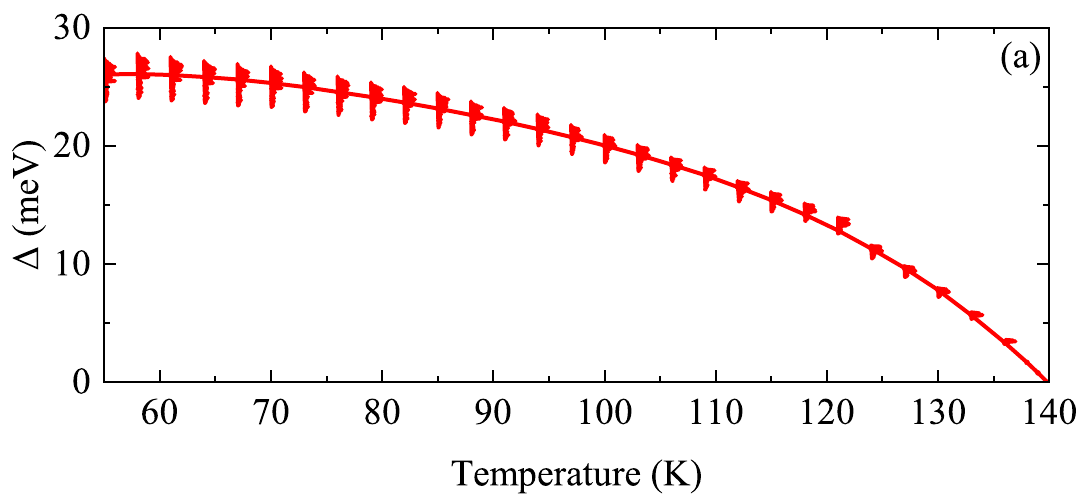}
\includegraphics[width=8.6cm]{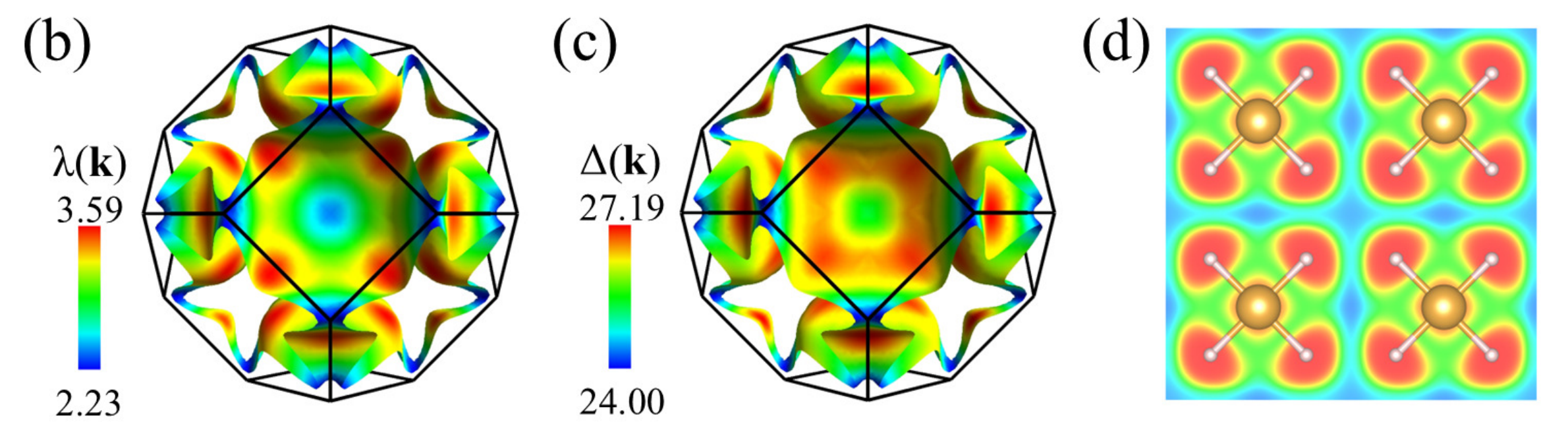}
\caption{(a) Normalized anisotropic superconducting gap $\Delta$ of Li$_2$AuH$_6$ under ambient pressure. The screening Coulomb potential $\mu^*$ is set to be 0.1 (b) Distribution of EPC $\lambda(\bf{k})$ on the Fermi surfaces. (c) Distribution of superconducting gap $\Delta(\bf{k})$ at 55 K on the Fermi surfaces. (d) Electron localization function of Li$_2$AuH$_6$. The yellow and pink atoms denote Au and H atoms, respectively.}
\label{fig:gap}
\end{figure}

In order to determine superconducting $T_c$ of Li$_2$AuH$_6$ under ambient pressure, we solve the anisotropic Eliashberg equations and the superconducting gap $\Delta$ is shown in Fig.~\ref{fig:gap}(a). The superconducting gap $\Delta$ is $\sim$ 26 meV at 55 K and eventually disappears at $\sim$ 140 K , which suggests that Li$_2$AuH$_6$ possesses a high superconducting $T_c$ $\sim$ 140 K. Figures~\ref{fig:gap}(b) and (c) show the anisotropic EPC and superconductivity of Li$_2$AuH$_6$. Relatively strong EPC electronic states exhibit a ring-like distribution upon the hemispherical Fermi surfaces. On the contrary, the electrons around the van Hove singularity contribute relatively weak EPC.

In Fig.~\ref{fig:gap}(d), we show the electron localization function (ELF) of the crystal plane where the Au and H atoms lie. It is clearly seen that almost all valence electrons localize around H atoms. This kind of charge density is easily expected to change dramatically under those previously mentioned strong EPC phonon modes. For example, the breathing mode of Au-H octahedron and $E_g$ mode at the $X$ point where Li and H atoms vibrate oppositely and get closer to each others [Fig.~\ref{fig:phonon}(d-f)]. Generally, metallic covalent bond is supposed to be beneficial to stability of crystal lattice and its metallization may induce strong EPC as well as high-$T_c$ superconductivity. However, in this situation of Li$_2$AuH$_6$, there are no covalent bonds between Au and H atoms while strong EPC and high $T_c$ still exist. As an intuitive comparison, Pd and H possess quite close electronegativity which suggests that electrons may exhibit more covalent properties than that in Au-H system. But the similar Li-Pd-H system~\cite{10.1063/5.0231618} is not dynamically stable under ambient pressure and even shows a lower predicted superconducting $T_c$ until a high pressure of 60 GPa than the ambient case of Li-Ag-H here. This may bring some helpful inspire regarding to search of potential conventional superconductors in the future.

\textit{Discussion and Conclusion.}
The establishment of Bardeen-Cooper-Schrieffer (BCS) theory~\cite{PhysRev.106.162} provides a theoretical support for understanding and searching conventional superconductors. And metallizing the $\sigma$-bonding electrons~\cite{PhysRevB.91.045132,2015-44-07-001} further proposes a possible way in finding high-$T_c$ superconductors, which has also achieved success in explaining superconductivity of MgB$_2$~\cite{WOS:000167194300040}. Specifically, the $\sigma$-bonds formed by the in-plane B-2$p_{x/y}$ electrons strongly couple with an in-plane vibration mode of boron atoms~\cite{PhysRevLett.86.5771}. Guiding by this, lots of theoretical works are focusing on finding materials with metallic $\sigma$-bands. B possesses close electronegativity to H, hence B-H system is regarded as a promising candidate. For example, KB$_2$H$_8$ is predicted to be a high-$T_c$ BCS superconductor due to the strong EPC between the metallic B-H $\sigma$-electrons and H-related phonon modes~\cite{PhysRevB.104.L100504}. However, our studies here find the Au-H system without metallic $\sigma$-electrons can also exhibit high-$T_c$ superconductivity, which leads us to realize that the strong coupling phonon that can cause a significant change of electronic density may be a more critical factor.

Hence, we propose a concept of BCS superconducting unit, which provides strong EPC phonon modes as superconducting pairing glue, just like the superconducting unit of CuO planes provides antiferromagnetic fluctuation as glue for Cooper pairing in cuprates. Furthermore, H octahedron with strong EPC phonon modes may be a potential BCS superconducting unit. One can try to design or synthesize more similar BCS superconducting units, such as H tetrahedron, etc., and stabilize them in crystal lattice to find high-$T_c$ BCS superconductors. In addition, considering the rising approaches of  high-throughput calculations, crystal structure prediction, and artificial intelligence~(AI), intercalating atoms directly in existing materials may introduce strongly coupled phonons into the system, which is a feasible way to investigate superconducting multicomponent compounds.

In this work, we first employed advanced AI technologies~\cite{han2024invdesflowaisearchengine} to identify candidate materials with high $T_c$, and then conducted high-precision DFT calculations for the candidate materials recommended by AI. This approach differs from the traditional high-throughput DFT screening of materials. High-throughput methods, limited by the computational cost of DFT, can only perform coarse calculations for all materials, which frequently leads to misleading results due to insufficient computational accuracy. In contrast, AI-based searching can more precisely narrow down the range of target materials, thereby allowing more computational resources (such as introducing anisotropic corrections, refined electron-phonon coupling analysis, and other advanced computational strategies) to be allocated to more valuable candidate materials.

In summary, we used our developed AI search engine, namely InvDesFlow~\cite{han2024invdesflowaisearchengine}, identified a possible superconducting hydride Li$_2$AuH$_6$. Our studies of thermodynamics suggest a feasible route of experimental synthesis. By performing the first-principles density-functional theory and superconducting calculations, we find that Li$_2$AuH$_6$ exhibits a superconducting transition temperature $\sim$ 140 K under ambient pressure. In addition to phonon modes of Au-H octahedrons that contribute strong EPC, vibrations of Li atoms also play a crucial role. Furthermore, we propose that searching those superconducting unit with strong EPC phonon modes may be an effective approach by combining with methods of high-throughput calculations, crystal structure prediction, and artificial intelligence for finding high-$T_c$ BCS superconductors.

\textit{Acknowledgments.} This work was financially supported by the National Natural Science Foundation of China (Grant No.62476278, No.12434009, and No.12204533). Z.Y.L. was also supported by the National Key R\&D Program of China (Grants No. 2024YFA1408601). Computational resources have been provided by the Physical Laboratory of High Performance Computing at Renmin University of China.

\bibliography{references}

\end{document}